\documentclass[12pt,thmsa]{article}

\begin{document}

\author{Preston Jones \\
%EndAName
Department of Mathematics and Physics\\
University of Louisiana at Monroe\\
Monroe, LA 71209-0575}
\title{A Dyad Theory of Hydrodynamics and Electrodynamics}
\date{January 15, 2007}
\maketitle

\begin{abstract}
The dyadic calculus is developed in a form suitable for the description of
physical relations in curved space. The 4-space equations of hydrodynamics
and electrodynamics are constructed using this dyadic calculus. As a
demonstration of the relationship between gravity and electrodynamics a time
varying metric is shown to generate electromagnetic radiation.

\end{abstract}

\section{Introduction}

In the early part of the last century the theory of general relativity was
developed by Einstein \cite{Einstein} using the component tensor calculus,
consistent with 4-space Riemann geometry. Today, this component tensor
calculus remains the most widely used method for the representation of
physics in 4-space. However, the component tensor calculus is not completely
suitable for the rigorous development of the mathematical formalism of
general relativity and the associated 4-space Riemann geometry. To place
general relativity and Riemann geometry on a rigorous mathematical
foundation recourse is made to the theory of differential forms \cite{MTW} 
\cite{Schutz80} \cite{Schutz90}. While differential forms can provide a
rigorous foundation for the theory of general relativity this formalism is
cumbersome or even useless in practical applications. The result is, that in
practice, both the component tensor calculus and differential forms must be
employed together in the theory of general relativity.

The goal here is to provide a bridge between the mathematically rigorous
differential forms and more practically useful component tensor calculus.
This will be achieved through the development of a middle ground in the
formalism of the dyadic calculus. The dyadic calculus has been discussed by
Goldstein \cite{Goldstein} and used to represent the tensor as a second rank
object in Euclidean 3-space. The formalism of the dyadic calculus was
extended by Luehr and Rosenbaum \cite{Luehr} to provide a description of
electrodynamics in Minkowski space using the traditional 3-vector electric
and magnetic fields. However, this restriction to Minkowski space and
3-vector fields is only correct for inertial observers in flat space. In
order to make the dyadic calculus a more useful tool the formalism must be
further extended to describe 4-vector relations in curved 4-space.

One of the greatest advantages of the dyadic calculus over the component
tensor calculus is the heuristic construction of the correct equations
describing a physical system. Using the dyadic calculus the equations are
always form invariant. That is the equations written in terms of 4-vectors
and their direct products are identical in all reference frames. This form
invariance insures that if the equations describing a system can be found
under the conditions of any specific reference frame the equations will be
the same for every reference frame. In an inertial frame and flat space the
physical relations can generally be written as differential operations on
3-space objects. By transforming these 3-space relations into the equivalent
4-space relations the resulting equations are then correct in all reference
frames and in particular where the curvature of space might be very unlike
flat space.

After developing the formal operations of the dyadic calculus, consistent
with general relativity and curved 4-space, this formalism will be used to
construct the hydrodynamic equations. These hydrodynamic equations, in
curved space, are already well represented in the literature and written in
the component tensor calculus \cite{Einstein} \cite{MTW} \cite{Schutz80} 
\cite{Schutz90} \cite{Tauber}. These equations are found using the dyadic
calculus as a demonstration of the heuristic construction of a physical
theory in the formalism of the dyadic calculus. The electrodynamic equations
are also found using this same heuristic construction. While the
electrodynamic equations are also well represented in the literature \cite
{MTW} \cite{Tauber} the resulting equations, using the dyadic calculus,
include an explicit representation of the electric and magnetic fields. This
explicit dependence on the fields is less well known and is consistent with
the electrodynamics of Ellis \cite{Ellis} \cite{Sonego}.

The value of the dyadic calculus is in part aesthetic. Physical theories
written in the dyadic calculus are highly symmetric. The dyadic calculus
shares all the symmetries of the component tensor calculus and differential
forms and is also symmetric with respect to covariant or contravariant
coordinate transformations. Unlike the component tensor calculus, the dyadic
calculus is completely symmetric with respect to flat space in the absence
of gravity and curved space in the presence of gravity. This final symmetry
leads to the expectation of a relationship between gravity and the electric
and magnetic fields in the equations of electrodynamics. This relationship
is demonstrated by considering the effect of a time varying metric on the
electric and magnetic fields in the equations of electrodynamics. The time
varying metric is shown to generate electromagnetic radiation. This gravity
induced electromagnetic radiation has not been previously predicted.

\section{Dyadic calculus}

The dyadic calculus, as presented here, is an extension of the ``intrinsic
tensor techniques'' of Luehr and Rosenbaum \cite{Luehr}. The ``intrinsic
tensor techniques'' provide an intuitive representation of electrodynamics
by expanding the differential operations of the vector calculus in Euclidean
space to Minkowski space. While the differential operations of the dyadic
calculus are consistent with the component tensor calculus, the dyadic
calculus is more restrictive in requiring the representation of all physical
relations in terms of 4-vectors. It is this restriction to 4-vectors that
retains the intuitive advantages of the ``intrinsic tensor techniques''
while also representing these relations in the curved space of general
relativity.

In the 3-space vector calculus geometric objects are represented as the
product of direction unit vectors, in a convenient coordinate system, and
the components as the magnitude of the object in that direction, 
\begin{equation}
\mathbf{A}\equiv A^{i}\mathbf{e}_{i}=A_{i}\mathbf{e}^{i};i=1,2,3.
\label{vector 1}
\end{equation}
\noindent The 3-vectors will be written in bold or with Latin indices. These
vectors are invariant in Euclidean space and a Galilean time transformation.
However, the Galilean time transformation is only an approximation of the
physically correct Lorentz time transformation. In order to construct
geometric objects that are invariant under a Lorentz time transformation a
fourth temporal direction must be included with the 3-space unit vectors, 
\begin{equation}
\underline{\mathbf{A}}\equiv A^{\alpha }\mathbf{e}_{\alpha }=A_{\alpha }%
\mathbf{e}^{\alpha };\alpha =0,1,2,3.  \label{vector 2}
\end{equation}

\noindent The 4-space objects will be represented in bold and underlined or
indicated by Greek indices. These 4-vectors are invariant in Minkowski
space, $A^{\alpha }\mathbf{e}_{\alpha }=A^{\alpha ^{\prime }}\mathbf{e}
_{\alpha ^{\prime }},$ where the primed and the unprimed represent the
components and basis vectors in two different Lorentz inertial frames (LIF).
The LIF is assumed to not be rotating or accelerating and to be far enough
away from any gravitational sources that the space can be considered flat.
More importantly, for the present purpose, is that these 4-space objects are
form invariant and represent the same physical phenomena in flat space of
special relativity and in the curved space of general relativity.

In retaining the basis vectors there is a close relation between
differential operations in the dyadic calculus and the vector calculus. This
is achieved by the construction of second rank objects from 4-vectors using
the definition of the direct product. This direct product produces a new
object the dyad that does not exist in vector calculus, 
\begin{equation}
\underline{\mathbf{A}}\underline{\mathbf{B}}\equiv \underline{\mathbf{A}}
\otimes \underline{\mathbf{B}}.  \label{Direct Product}
\end{equation}
This new object is the juxtaposition of two 4-vectors and is sometimes
referred to as a second rank object. The order of the 4-vectors is important
to the definition of the object and in the differential operations on the
object. The objects of the dyadic calculus are all 4-vectors and the direct
products of 4-vectors and this preserves the contributions from derivatives
of the base vectors $\mathbf{e}_{\beta }$ \cite{Foster}, 
\begin{equation}
\mathbf{e}_{\gamma }\Gamma _{\beta \alpha }^{\gamma }\equiv \frac{\partial }{
\partial x^{\alpha }}\mathbf{e}_{\beta },  \label{Christofel}
\end{equation}
where the $\Gamma _{\beta \alpha }^{\gamma }$\ are the connection
coefficients. The definition of the derivatives of the basis vectors
provides a formal connection between the differential operations of the
dyadic calculus and the covariant derivative of the component tensor
calculus.

The scalar product in 4-space is very similar to the vector calculus
operation $\underline{\mathbf{A}}\cdot \underline{\mathbf{B}}\equiv
A^{\alpha }B^{\beta }g_{\alpha \beta }=A_{\alpha }B_{\beta }g^{\alpha \beta
} $. The most significant difference is the negative signature of the metric
in 4-space. This requires that a sign convention be adopted for either a
negative space interval or negative time interval. Here the time interval is
taken to be negative and in a LIF $g_{00}=-1.$ The metric components are the
scalar products of the basis vectors and the scalar product of the time
basis, in a LIF, can be expressed as $g_{00}=\mathbf{e}_{0}\cdot \mathbf{e}
_{0}=\mathbf{e}^{0}\cdot \mathbf{e}^{0}=-\mathbf{e}_{0}\cdot \mathbf{e}^{0}$%
. In this context the scalar products of the remaining base vectors form the
Kronecker delta, $\mathbf{e}_{i}\cdot \mathbf{e}^{j}=\delta _{i}^{j}$.
Differential operations are introduced by defining a 4-space differential
operator $\underline{\mathbf{\nabla }}\equiv \mathbf{e}^{\delta }\frac{
\partial }{\partial x^{\delta }}=\mathbf{e}_{\delta }\frac{\partial }{
\partial x_{\delta }},$ where in a LIF for example $x^{0}=-x_{0}=-ct$ and $%
\underline{\mathbf{\nabla }}=-\frac{1}{c}\mathbf{e}_{0}\frac{\partial }{%
\partial t}+\mathbf{\nabla }=\frac{1}{c}\mathbf{e}^{0}\frac{\partial }{%
\partial t}+\mathbf{\nabla }.$ The divergence of a 4-vector in curvilinear
coordinates is defined as a natural extension of the vector calculus
expansion,

\begin{equation}
\underline{\mathbf{\nabla }}\cdot \underline{\mathbf{A}}\equiv \mathbf{e}
_{\delta }\cdot \frac{\partial }{\partial x_{\delta }}A_{\alpha }\mathbf{e}
^{\alpha }=\mathbf{e}^{\delta }\cdot \frac{\partial }{\partial x^{\delta }}
A^{\alpha }\mathbf{e}_{\alpha }.  \label{Divergence}
\end{equation}
\noindent The divergence of the direct product of two vectors is defined in
a similar fashion, 
\begin{equation}
\underline{\mathbf{\nabla }}\cdot (\underline{\mathbf{A}}\underline{\mathbf{B%
}})\equiv (\underline{\mathbf{A}}\cdot \underline{\mathbf{\nabla }})%
\underline{\mathbf{B}}+\underline{\mathbf{B}}(\underline{\mathbf{\nabla }}
\cdot \underline{\mathbf{A}}).  \label{2 Dot}
\end{equation}

Unlike the scalar product, the vector product of vector calculus does not
have an obvious 4-space analogue. To define a 4-space analogue to the vector
product two further operations, the wedge product and the dual, are
required. The wedge product is the anti-symmetrization of the direct product
of two vectors,

\begin{equation}
\underline{\mathbf{A}}\wedge \underline{\mathbf{B}}\equiv \underline{\mathbf{%
\ A}}\otimes \underline{\mathbf{B}}-\underline{\mathbf{B}}\otimes \underline{%
\mathbf{A}}.  \label{Wedge}
\end{equation}
The dual maps the 4-space object into the ``dual'' of the object, 
\begin{equation}
dual(\underline{\mathbf{A}}\underline{\mathbf{B}})\equiv \frac{1}{2}
e^{\alpha \beta \gamma \delta }A_{\alpha }B_{\beta }\mathbf{e}_{\gamma }%
\mathbf{e}_{\delta }=\frac{1}{2}e_{\alpha \beta \gamma \delta }A^{\alpha
}B^{\beta }\mathbf{e}^{\gamma }\mathbf{e}^{\delta }.  \label{Dual}
\end{equation}
The Levi-Civita tensor, 
\begin{equation}
e^{\alpha \beta \gamma \delta }\equiv -\frac{1}{\sqrt{-g}}E^{\alpha \beta
\gamma \delta }=\sqrt{-g}E_{\alpha \beta \gamma \delta }=e_{\alpha \beta
\gamma \delta },  \label{Levi-Civita}
\end{equation}
is defined in terms of the permutations symbol $E^{\alpha \beta \gamma
\delta }$ which is zero if any indices are repeated, one for even
permutations of $0,1,2,3$ and negative one for odd permutations. The
coefficient $g\equiv \det \left( \mathbf{e}_{\alpha }\cdot \mathbf{e}_{\beta
}\right) $ is the determinant of the metric components.

\section{Hydrodynamics}

Satisfying the requirement of form invariance, requiring that the equations
are the same in any reference frame, is where the formalism of the component
tensor calculus and the dyadic calculus most differ. Form invariance is
satisfied in the component tensor calculus by changing from the partial
derivative in the flat space of inertial reference frames to the covariant
derivative in more general curved spaces. In the dyadic calculus, form
invariance is insured by calculating the derivatives of the basis vectors in
the partial derivatives. As a demonstration of the application of form
invariance, using the dyadic calculus, the equations of motion for a perfect
fluid will be developed heuristically by determining these equations in flat
space. The form invariance of the 4-space objects insures that these objects
are exactly the same in curved space. Using the dyadic calculus the correct
equations in flat space and special relativity are the same equations in
curved space and general relativity.

Assume that the volume element of a perfect fluid, in a LIF, has nonzero
temperature or pressure $P$, 3-velocity $\mathbf{u}$\textbf{,} and
4-velocity $\underline{\mathbf{u}}=\gamma \mathbf{e}_{0}+\gamma u^{c}\mathbf{%
e}_{c}$ in a system of units with the speed of light $c=1$. Define the
finite temperature mass current density of this fluid as

\begin{equation}
\underline{\Omega }=\gamma \tilde{\sigma}\mathbf{e}_{0}+\gamma \tilde{\sigma}%
u^{a}\mathbf{e}_{a},  \label{4-momentum}
\end{equation}

\noindent where $\tilde{\sigma}=\left( P+\sigma \right) $ is the local
mass-energy density, $\sigma $ is the local rest mass density, and $P$ is
the pressure. If the particles of the fluid could be made stationary,
relative to one another, each particle would have exactly the same motion in
a chosen frame. While these particles would have a different motion in some
other frame the particles relative motion, in that frame, will still be the
zero. This permits the definition of a second physical object, the zero
temperature mass current density, which in the LIF is

\begin{equation}
\underline{\mathbf{p}}=\gamma \sigma \mathbf{e}_{0}+\gamma \sigma u^{b}%
\mathbf{e}_{b}.  \label{0 temperature}
\end{equation}

\noindent It is now only necessary to show that, in the low energy limit,
the continuity and Euler equations for a perfect fluid in a LIF and written
as 4-vector relations are

\begin{equation}
\underline{\mathbf{\nabla }}\left( \underline{\mathbf{u}}\cdot \underline{%
\Omega }\right) -\underline{\mathbf{\nabla }}(\underline{\mathbf{u}}\cdot 
\underline{\mathbf{p}})=\underline{\mathbf{\nabla }}\cdot \left( \underline{%
\mathbf{u}}\underline{\Omega }\right) .  \label{Hydro eq}
\end{equation}

\noindent Form invariance insures that this expression is correct in any
reference frame and in curved space.

As an illustration of the similarity between the dyadic calculus and the
more familiar vector calculus the equation for a perfect fluid will be
expanded in some detail. Using the definition of the divergence of the
direct product, the right hand side can be expanded as 
\begin{equation}
\underline{\mathbf{\nabla }}\cdot \left( \underline{\mathbf{u}}\underline{%
\Omega }\right) =\left( \underline{\mathbf{u}}\cdot \underline{\mathbf{
\nabla }}\right) \underline{\Omega }+\underline{\Omega }\left( \underline{%
\mathbf{\nabla }}\cdot \underline{\mathbf{u}}\right) .  \label{diverge}
\end{equation}

\noindent Assuming a LIF the equation can be expanded and assuming the low
energy limit terms proportional to $\frac{1}{c^{2}}$ can be dropped,

\begin{eqnarray}
-\frac{\partial }{\partial x^{a}}P\mathbf{e}_{a} &=&\left( \frac{\partial }{%
\partial t}\sigma +u^{d}\frac{\partial }{\partial x^{d}}\sigma +\sigma \frac{
\partial }{\partial x^{d}}u^{d}\right) \mathbf{e}_{0}  \nonumber \\
&&+\left( \frac{\partial }{\partial t}\sigma u^{a}+\sigma u^{a}\frac{%
\partial }{\partial x^{d}}u^{d}+u^{d}\frac{\partial }{\partial x^{d}}\sigma
u^{a}\right) \mathbf{e}_{a}.
\end{eqnarray}

\noindent Note that for low energies $\gamma =\left( 1-\frac{u^{2}}{c^{2}}%
\right) ^{-\frac{1}{2}}\simeq 1$ and the derivatives of $\gamma $ are $\frac{%
\partial \gamma }{\partial x^{\alpha }}=\frac{u}{c^{2}}\gamma ^{3}\frac{%
\partial u}{\partial x^{\alpha }}\simeq 0$. Equating the terms for the space
part, this expression can also be written as a 3-space dyadic equation,

\begin{equation}
\frac{\partial }{\partial t}\sigma \mathbf{u}+\mathbf{\nabla }\cdot \left(
\sigma \mathbf{uu}\right) =-\mathbf{\nabla }P.  \label{3 dyad}
\end{equation}

\noindent Collecting terms for the time part,

\begin{equation}
\frac{\partial }{\partial t}\sigma +\mathbf{\nabla }\cdot \left( \sigma 
\mathbf{u}\right) =0.  \label{low energy}
\end{equation}

\noindent These are the Euler and continuity equations for a perfect fluid
in a LIF and the low energy limit.

\section{Electrodynamics}

The dyadic calculus, as it is presented here, makes it possible to write a
form invariant expression of the Maxwell equations in terms of the direct
products between a 4-velocity and the 4-vector electric $\underline{\mathbf{E%
}}$ and magnetic $\underline{\mathbf{B}}$ fields. The 4-velocity $\underline{%
\mathbf{u}}$ is the velocity of the volume element where the electric and
magnetic fields are defined. The Maxwell equations were previously written
in a similar form, as direct products between the 4-velocity and 4-space
fields, by Ellis \cite{Ellis}\textit{.} While Ellis writes the equations of
electrodynamics using the component tensor calculus the 4-space objects are
the same.

The inhomogeneous equations of electrodynamics are written in terms of the
wedge product of the 4-velocity $\underline{\mathbf{u}}$ and the 4-vector $%
\underline{\mathbf{E}}$ field and the dual of this wedge product with the
4-vector $\underline{\mathbf{B}}$ field,

\begin{equation}
\underline{\mathbf{\nabla }}\cdot (\underline{\mathbf{u}}\wedge \underline{%
\mathbf{E}})+\underline{\mathbf{\nabla }}\cdot dual(\underline{\mathbf{u}}
\wedge \underline{\mathbf{B}})=-4\pi \underline{\mathbf{J}}.
\label{Nonhomogenious}
\end{equation}
The homogeneous equations are written as the wedge product with the 4-vector 
$\underline{\mathbf{B}}$ field and the dual of the wedge product with the
4-vector $\underline{\mathbf{E}}$ field,

\begin{equation}
\underline{\mathbf{\nabla }}\cdot (\underline{\mathbf{u}}\wedge \underline{%
\mathbf{B}})=\underline{\mathbf{\nabla }}\cdot dual(\underline{\mathbf{u}}
\wedge \underline{\mathbf{E}}).  \label{Homgenious}
\end{equation}
In this form the physical content of the equations are independent of the
observer and the reference frame and depend only on the source terms $%
\underline{\mathbf{J}}$.

Since these equations are form invariant it will suffice to show that the
equations are correct for a LIF and flat space to demonstrate that the
equations are correct for curved space as well. In a LIF take the 3-velocity
as zero,$\mathbf{\ u=}0$ and $\underline{\mathbf{u}}=\mathbf{e}_{0}=-\mathbf{%
e}^{0}$. A system of units is assumed where the speed of light $c=1$. With
this definition the source terms are $\underline{\mathbf{J}}=\rho \mathbf{e}%
_{0}+\mathbf{J}=-\rho \mathbf{e}^{0}+\mathbf{J}.$ The fields are similarly
expressed as a sum of a time and space part. Substituting these expressions
into the general form of the Maxwell equations and collecting space and time
terms reduces to the expected form of the equations for an observer at rest
in a LIF. The connection between 4-space and 3-space objects and operators
is facilitated by recognizing the relation between the Levi-Civita tensors
in a LIF and assuming constant base vectors, $e^{0123}=-e^{123}$. This
relation leads to an expression for the negative of the curl in terms of the
dual,

\begin{equation}
\underline{\mathbf{\nabla }}\cdot dual(\mathbf{e}_{0}\wedge \mathbf{A})=-%
\mathbf{\nabla \times A.}
\end{equation}

\noindent Expanding the inhomogeneous equations in a LIF,

\begin{equation}
\frac{\partial \mathbf{E}}{\partial t}-\left( \nabla \cdot \mathbf{E}\right) 
\mathbf{e}_{0}-\nabla \times \mathbf{B}=-4\pi \left( \mathbf{J}+\rho \mathbf{%
e}_{0}\right) \mathbf{.}  \label{Inhomo}
\end{equation}

\noindent Expanding the homogeneous equations in a LIF,

\begin{equation}
\left( \nabla \cdot \mathbf{B}\right) \mathbf{e}_{0}-\frac{\partial \mathbf{B%
}}{\partial t}=\nabla \times \mathbf{E.}  \label{Homo}
\end{equation}

\noindent These equations are correct for flat space and the form invariance
of the dyads insures that these equations are correct for all reference
frames and curved spaces.

\section{Gravity and electromagnetic radiation}

Having established the form invariant equations for electrodynamics, it is
evident that the magnitude of the 4-space electric and magnetic fields must
be related to the local curvature of space and the gravitational sources
associated with that curvature. As a practical example of this connection,
between gravity and electromagnetism, the effect of a time varying metric on
the electric and magnetic fields will be considered. Assuming time dependent
gravity, in some region of space, the metric is written as $g_{00}=-\left(
1+h\left( t\right) \right) ,$ and, $g_{ii}\simeq 1\mathbf{,}$ $g_{ij}=0$ if $%
i\neq j$. Assume that the field is weak, $h\prec \prec 1,$ and that the
spacial derivatives of the field are small compared to the time derivative, $%
\frac{\partial }{\partial x^{i}}h\prec \prec \frac{\partial }{\partial t}h$.
This metric is similar to the metric of the usual weak field limit, except
that the field here is time dependent and independent of position. Also
assume that there are no electromagnetic sources, $\underline{\mathbf{J}}=0.$
The choice of reference frames is made where the volume element is
stationary,\textit{\ }$\underline{\mathbf{u}}=\mathbf{e}_{0}=-\mathbf{e}^{0}$%
\textit{.}

The time dependent gravity and the equations of electrodynamics are
operationally related by the time dependence of the temporal basis vector.
The effect of the time varying metric on the equations of electrodynamics
can be calculated explicitly by equating the time derivatives of the basis
vectors with the time derivative of the metric component. The only non zero
contribution is from the temporal unit basis in the first term on the left
hand side of the ``inhomogeneous equations'', $\left( -\mathbf{e}_{0}\cdot 
\frac{\partial }{\partial t}\mathbf{e}_{0}\mathbf{E}\right) =\frac{\partial 
}{\partial t}\mathbf{E-E}\left( \mathbf{e}_{0}\cdot \frac{\partial }{
\partial t}\mathbf{e}_{0}\right) $. This last expression, in parentheses, on
the right hand side can be rewritten in terms of the time variation in
gravity, $\left( \frac{\partial }{\partial t}e_{0}\right) \cdot e_{0}=\frac{1%
}{2}\frac{\partial }{\partial t}g_{00}=-\frac{1}{2}\frac{\partial }{\partial
t}h$\textit{.} The ``inhomogeneous equations'' can then be expanded,
retaining the time dependence in the metric and taking the scalar product
with the electric field as,

\begin{equation}
-\mathbf{E}\cdot \left( \nabla \mathbf{\times B}\right) \mathbf{+E}\cdot 
\frac{\partial \mathbf{E}}{\partial t}=-\frac{1}{2}E^{2}\frac{\partial h}{%
\partial t}.  \label{E dot}
\end{equation}

\noindent The homogeneous equations are similarly expanded,

\begin{equation}
\mathbf{B}\cdot \left( \nabla \times \mathbf{E}\right) \mathbf{+B}\cdot 
\frac{\partial \mathbf{B}}{\partial t}=-\frac{1}{2}B^{2}\frac{\partial h}{%
\partial t}.  \label{B dot}
\end{equation}

\noindent Adding the equations and dividing by $\frac{1}{4\pi },$

\begin{equation}
\frac{1}{4\pi }\nabla \cdot \left( \mathbf{E\times B}\right) =-\frac{1}{8\pi 
}\left( B^{2}+E^{2}\right) \frac{\partial h}{\partial t}-\frac{1}{8\pi }%
\frac{\partial }{\partial t}\left( B^{2}+E^{2}\right) ,  \label{Divide}
\end{equation}

\noindent where the identity $\nabla \cdot \left( \mathbf{a\times b}\right) =%
\mathbf{b}\cdot \left( \nabla \times \mathbf{a}\right) -\mathbf{a}\cdot
\left( \nabla \mathbf{\times b}\right) $ has been used to simplify the left
hand side. Substituting the local energy density $U=\frac{1}{8\pi }\left(
E^{2}+B^{2}\right) $ and the Poynting vector $\mathbf{S=\frac{1}{4\pi }%
E\times B}$, 
\begin{equation}
\nabla \cdot \mathbf{S}=\frac{\partial }{\partial t}\left( Ug_{00}\right) .
\label{Poynting}
\end{equation}

\noindent This demonstrates that time variations in gravity will generate
electromagnetic radiation.

\section{Conclusion}

The dyadic calculus, in the present form, offers a middle ground between the
computational utility of the component tensor calculus and the mathematical
rigor of differential forms. As a demonstration of the heuristic
construction of a physical theory, in curved space, the equations of
hydrodynamics and electrodynamics were developed using the dyadic calculus.
Recognizing the connection between gravity and the electric and magnetic
fields, in the equations of electrodynamics, gravity was shown to be a
potential source of electromagnetic radiation.

\section*{Acknowledgment}

The author would like to thank Richard Morris, Marcos Rosenbaum, and Peter
Winkler for their helpful discussions of dyads and general
relativity.\pagebreak

\end{document}